\documentclass{article}                     

\usepackage{graphicx}
\usepackage{color}

\begin{document}

\title{Interaction of free-floating planets with a star-planet
pair}

\author{Harry Varvoglis, Vasiliki Sgardeli and Kleomenis Tsiganis
\\\
              Department of Physics, University of Thessaloniki \\ GR-54124 Thessaloniki, Greece \\
              E-Mail: varvogli@physics.auth.gr}

\maketitle

\begin{abstract}
The recent discovery of free-floating planets and their theoretical
interpretation as celestial bodies, either condensed independently
or ejected from parent stars in tight clusters, introduced an
intriguing possibility. Namely, that exoplanets are not condensed
from the protoplanetary disk of their parent star. In this novel
scenario a free-floating planet interacts with an already existing
planetary system, created in a tight cluster, and is captured as
a new planet. In the present work we study this interaction
process by integrating trajectories of planet-sized bodies, which
encounter a binary system consisting of a Jupiter-sized planet
revolving around a Sun-like star. To simplify the problem we assume
coplanar orbits for the bound and the free-floating planet and an initially
parabolic orbit for the free-floating planet. By calculating the
uncertainty exponent, a quantity that measures the dependence of the
final state of the system on small changes of the initial
conditions, we show that the interaction process is a fractal
classical scattering. The uncertainty exponent is in the range $(0.2-0.3)$ and is a decreasing function of time. In this way we see that the statistical approach we follow
to tackle the problem is justified. The possible final outcomes of
this interaction are only four, namely flyby, planet exchange,
capture or disruption.  We give the probability of each outcome as a
function of the incoming planet's mass. We find that the probability of exchange or capture (in prograde as well as retrograde orbits and for very long times) is non-negligible, a fact that might explain the
possible future observations of planetary systems with orbits that are either retrograde or tight and highly eccentric.

\end{abstract}

\section{Introduction}
\label{intro} The three-body problem has been a topic of active
research since the seventeenth century, initially most of the effort
focusing on bound solutions and their long time stability. This
problem was ``solved" through the work of
(a) Poincar\'e 1893, who showed that the general solutions of even
the simplest variant of the problem (apart from highly symmetric configurations), the restricted planar circular
three-body problem, cannot be written in the form of convergent
series, and (b) Marchal and Bozis (1982), who showed that in the
general three body problem zero velocity curves do not exist, so
that one of the bodies can always escape to infinity. However the
``inverse" problem, i.e. the interaction of a body coming from
``infinity" with a binary pair, has attracted less
attention. From the astrophysical point of view, there are three
distinctive special forms of this problem: (a) the binary members
and the incoming body have similar masses (interaction of a
``passing" star with a binary star), (b) the binary consists of a
massive and a small body and the incoming body is massive (a passing
star interacting with a planetary system) and (c) the binary
consists of a massive and a small body and the incoming body is
small (a passing planet interacting with a planetary system). The
first two forms have already been discussed in the literature (e.g.
see Boyd \& MacMillan, 1993; Mikkola, 1994; Donnison, 2008 for the first,
Astakhov and Farrelly, 2004 for the second). However the third one
has attracted limited interest up to now (e.g. see Donnison, 2006),
probably because planets were considered always to be bodies
revolving about stars. The discovery of free floating planets (FFPs) a few
years ago changed this picture.

The initial report by Zapatero Osorio et al. (2000) on the
observation of free-floating planets with masses $m_J<m<3m_J$ (where $m_J$ is Jupiter's mass) was regarded initially with
scepticism, since it was not consistent with the generally accepted
definition of a planet. However subsequent microlensing observations (Sumi et al., 2011)
showed that celestial bodies with masses $m \approx m_J$ really do exist in the Milky Way, and for that in numbers exceeding those of its stars, making the interaction of FFPs
with already existing planetary systems a noteworthy process. In any
case, recent publications (e.g. see Adams {\it et al., 2006; Malmberg {it et al.}, 2007})
have shown that the probability of such an interaction is
non-trivial within clusters of newborn stars, where a planet ejected
from a young planetary system (or condensed independently) may
remain within the cluster for long time intervals. In the present
paper we study numerically the simplest possible model of such an
interaction, namely a bound planet (BP) on circular orbit around a star,
scattering a FFP approaching from ``infinity" on a coplanar orbit.

\begin{figure}
\begin{center}
  \includegraphics[width=11cm] {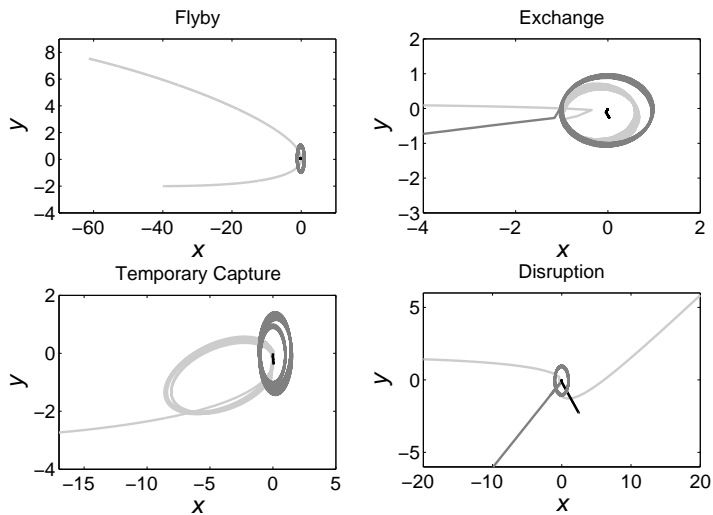}
\end{center}
\caption{Characteristic orbits of the scattering between a star - planet system and a free-floating planet (FFP)
(FFP: light gray, bound planet (BP): dark gray, star: black). {\it Flyby}: a
simple deflection of the FFP, {\it exchange}: the FFP displaces
the BP and takes its place orbiting the star, {\it temporary capture}: the two bodies
are temporarily bound to the star, {\it disruption}: the system
breaks up and the three bodies move to infinity. Note that disruption is not possible
when the system's center of mass energy is negative, which is the case in our numerical experiments. }
\label{fig:0}       
\end{figure}

The interaction of a FFP with an existing
star-planet binary system has only four possible outcome states: the
incoming body (a) goes to infinity, probably exciting the planet (flyby), (b) replaces in orbit
around the star the BP, which in turn goes to infinity
(exchange) (c) is captured by the star-planet binary system, forming
a two-planet planetary system (temporary capture) or (d) the system is
disrupted (disruption) (Fig. \ref{fig:0}). If the initial total energy of the system
is negative, as in our numerical experiments, the fourth outcome is
ruled out and we are left with only three possible outcomes.

From the purely dynamical point of view, however, the situation is
more complicated. The interaction can be distinguished into
``single-" and ``multiple-encounter" events. In a single-encounter
event the FFP interacts only briefly with the BP (e.g. for a time interval less than one unperturbed period of the BP)
while in a multiple encounter event the FFP enters in orbit around the
star and interacts for an extended period of time with the BP (typically much more than one unperturbed period). Therefore a
single encounter event can result only in flyby or exchange,
which may be thought of as ``direct" outcomes. To the contrary, a
multiple encounter event (temporary capture) is not a ``definitive" outcome, since
after some time interval one of the two planets may escape to
infinity (e.g. see Marchal \& Bozis, 1982). In other words, a
temporary capture may ultimately decay to a flyby or an exchange, which, in
this sense, may be considered as ``indirect" outcomes. Therefore
finally a more comprehensive classification of the possible outcomes
may read as follows: flyby (direct and indirect), exchange (direct
and indirect) and ``long-time" capture. It is not possible to
guarantee a permanent capture since, as we mentioned already, in the general three-body
problem closed zero-velocity curves do not exist (except from cases that are Hill stable). But, as we show in section 4, the statistical analysis of our results indicates that a
percentage of temporary captures end up in ``deep" captures, in the sense
that the resulting two-planet system may remain stable for very
long times. In fact, if a fourth body exists nearby or other forces (e.g. tides) are taken into account, the 3-body system might be stabilized.

The paper is organized as follows. In the next section we describe
the setting of our numerical experiments. In section 3 we
show that the scattering process is a fractal one, we calculate
the uncertainty exponent and we give the probability of each one of
the possible outcomes of the scattering. In section 4 we
present our main results, concerning (a) the dependence of the
scattering outcomes on the mass of the incoming FFP, (b) the physics
behind capture and (c) the orbital distribution (in the cases of
capture and exchange) as a function of the mass and the impact parameter of the
FFP. In section 5 we deal with the decay of temporary captures and in section 6 we find the initial conditions that lead to prograde or retrograde orbits. Finally in section 7 we discuss our results.

\section{Numerical experiments setup}
\label{sec:1}
\subsection{Initial configuration}
\label{sec:2} 
\begin{figure}
\begin{center}
  \includegraphics[width=11cm] {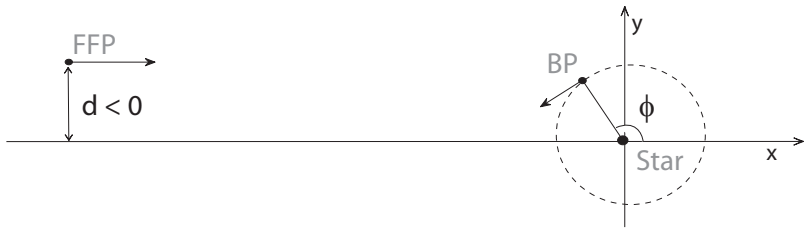}
\end{center}
\caption{Initial configuration of the scattering between a
star--planet system and a FFP. The figure shows the initial position of the three bodies
with respect to an inertial reference frame $Oxy$. The BP is initially on a circular orbit
around the star and the FFP is coming with parabolic co-planar velocity from a fixed distance,
sufficient to ensure that it is not initially interacting strongly with the planetary system.
The initial state of the system is determined by the impact parameter, $d$, of the FFP and the initial phase, $\phi$, of the BP.
Note that $d$ is negative for retrogade orbits ($y>0$) and positive for prograde orbits ($y<0$).}
\label{fig:1}       
\end{figure}

We perform numerical experiments of the three body scattering
between a star--planet system and an incoming FFP. The three bodies are considered as point
masses moving on coplanar orbits and are: a star of mass $M$,
a BP of mass $m_J$ in orbit around the
star, and the FFP with mass $m$ similar to that of the BP,
incoming from infinity. The initial configuration of the scattering
experiment is shown in Fig. \ref{fig:1}.
The star is initially placed at rest at the origin of the inertial reference frame $0xy$ and the BP
is placed at a distance $r_0$ with initial velocity, $v_0$, corresponding to a circular
orbit around the star. The FFP is coming from a fixed distance of
40$r_0$ with impact parameter $d$ and initial parabolic velocity $v$
with respect to the star. This distance is sufficient to ensure that
the FFP and the original planetary system are not initially interacting strongly. The
initial velocity, $v$, of the FFP is estimated as a function of $d$
through the relationship:

\begin{equation}
\label{eq1} E = \frac{1}{2}\mu \upsilon ^2 - G\frac{mM}{\sqrt {d^2 +
\left( {40r_0 } \right)^2} } = 0
\end{equation}

\noindent where $\mu = (mM)/(m+M)$. Keeping all other parameters fixed, the initial state of the system
is determined by the impact parameter of the FFP, $d$, and the
initial phase of the BP, $\phi$. Note that $d$ can take both positive and
negative values. By convention, we assign a negative value to $d$ when
the angular momentums of the BP and the FFP have different
signs ($y>0$ in Fig. \ref{fig:1}) and positive otherwise ($y<0$ in Fig.
\ref{fig:1}). The initial values of the impact
parameter are restricted in the interval $-7r_0 < d < 7r_0$, where
exchange and temporary capture are mainly expected to occur (see e.g.
Donnison, 1984a,b). Within the configuration of Fig. 2 the total energy of the
system at the center of mass is always negative, although it depends
weakly on $\phi$ and $d$.

It should be noted that the (dimensionless) interval of impact parameters chosen here is the one that yields dynamically interesting results. FFPs passing at a larger distance have minor effects on the BP. That is why the most promising site for this kind of interaction is in dispersing clusters (see, e.g., Perets \& Kouwenhoven, 2012). But the {\it dimensional} value of impact parameters in AUs depends on $r_0$. If one is taking as example our solar system, then he could consider as ``bound planet" (BP) anyone of the giant planets. When considering Jupiter as BP the impact parameter is 35 AU, while when considering Neptune it is 210 AU.

\subsection{Equations of motion and integration}
Our dynamical system is an autonomous one with six degrees of freedom and
is described by the Hamiltonian:

\begin{equation}
\label{eq2} H = \frac{\mathbf{\left|p_1\right|}^2 }{2m} + \frac{\mathbf{\left|p_2\right|}^2 }{2M} + \frac{\mathbf{\left|p_3\right|}^2
}{2m_J } - G\frac{mM}{\left| {\mathbf {r}_1 - \mathbf {r}_2 } \right|} -
G\frac{m_J M}{\left| {\mathbf {r}_3 - \mathbf {r}_2 } \right|} -
G\frac{mm_J }{\left| {\mathbf {r}_1 - \mathbf {r}_3 } \right|},
\end{equation}

\noindent with $\mathbf {r}_1$, $\mathbf {r}_2$ and $\mathbf {r}_3$ the positions of
the FFP, the star and the BP with respect to the inertial reference frame $x,y$ and $\mathbf{p}_1$, $\mathbf{p}_2$ and $\mathbf{p}_3$ the corresponding momenta.



\noindent The initial position and momentum of the three bodies are:
$x=0$, $y=0$, $p_{x}=0$, $p_{y}=0$ for the star,
$x=r_0 \cos \phi$, $y=r_0 \sin \phi$, $p_x=-mv_0 \sin \phi$, $p_y= mv_0 \cos \phi$ for the BP
and
$x=-40r_0$, $y=-d$, $p_x=m_J v$, $p_y=0$ for the FFP.

Throughout the numerical experiments we fix the mass ratio of the star - planet system equal to the mass ratio of the Sun - Jupiter system and we set the radius of the planet's orbit ($r_0$) equal to the semi-major axis of Jupiter. We use a system of units where $G=1$, $M=1$, $r_0=1$. It follows that the mass,
circular velocity and orbital period of the BP are
$m_J=0.000964583 M$, $v_0=1$ and $T=2 \pi$ respectively.
The initial velocity of the FFP is determined through eq. (\ref{eq1})
as a function of $d$ and $m$ and is approximately equal to
$v=0.2236 v_0$. The equations of motion (\ref{eq2}) are integrated
through a Rosenbrock scheme, which is optimized for stiff ordinary
differential equations (Rosenbrock, 1963; Gear, 1971).

\subsection{Final state -- exit tests}
Due to energy exchange between the three bodies the system may end up in
four qualitatively different states, depending on the total energy
of the system at the center of mass. In particular there are three
types of unbound motion. These are: \textit{flyby}, where the
FFP is deflected from the planetary system leaving it bound,
\textit{exchange}, where the FFP displaces the BP and
takes its place orbiting the star and \textit{disruption}, where the
planetary system is destroyed and the three bodies move
independently. Disruption is possible only when the total energy of
the system is zero or positive. Bound motion between the three
bodies is called \textit{temporary capture} and can occur only when the
total energy of the system is negative. At temporary capture, the FFP
and the BP are temporarily bound to the star. Again, due to
energy exchange between the bodies at every close approach, one of
the bodies (BP or FFP) can gain enough energy and escape.
So, temporary capture can result in \textit{indirect flyby} (escape of the
FFP) or \textit{indirect exchange} (escape of the BP).
But it is not known {\it a priori} if and when a
temporary capture will ``break up" and end in indirect flyby or
exchange.

The state of the system at a given time after scattering is a function of the
initial configuration, which is determined by the parameters $\phi$ and $d$, since we keep all other parameters fixed. We have to note that the center-of-mass energy of the system is a weak function of $\phi$ and $d$, but it is always negative allowing for bound motion (temporary capture) and ruling out disruption. Our purpose is to map the initial conditions ($\phi,d$) to the final states of the system
(flyby, exchange or temporary capture). In order to do this, we choose 500 values of $\phi$ and 700 values of $d$ uniformly distributed in the intervals $0 < \phi < 1$ ($\phi$ in cycles) and $-7r_0 < d < 7r_0$ respectively. $|d_{max}|=7r_0$ is the maximum impact parameter for which a direct exchange can occur and may be estimated, as a function of the FFP's mass, using known analytical approximations (e.g. see Donnison, 1984a). The equations of motion are integrated for each initial conditions set
(\textit{$\phi $},$ d)$ until a certain time $t_{test}$. At that time a test is made to decide the final state of
the system. However, the exit test cannot distinguish between direct and indirect, flyby or exchange, since the three final states are identified only on the basis of the relative energy of the bodies in
pairs at time $t_{test}$. If $E_{12}$, $E_{23}$, $E_{13}$ are the
center-of-mass energies of the pairs FFP-star, BP-star and
FFP-BP, they correspond to the final states as follows:

\begin{itemize}
  \item a. Flyby: $E_{12} > 0, E_{23} < 0, E_{13} > 0$
  \item b. Exchange: $E_{12} < 0, E_{23} > 0, E_{13} > 0$
  \item c. Temporary capture: $E_{12} < 0, E_{23} < 0, E_{13} > 0$
\end{itemize}

\begin{figure}
\begin{center}
  \includegraphics[width=11cm] {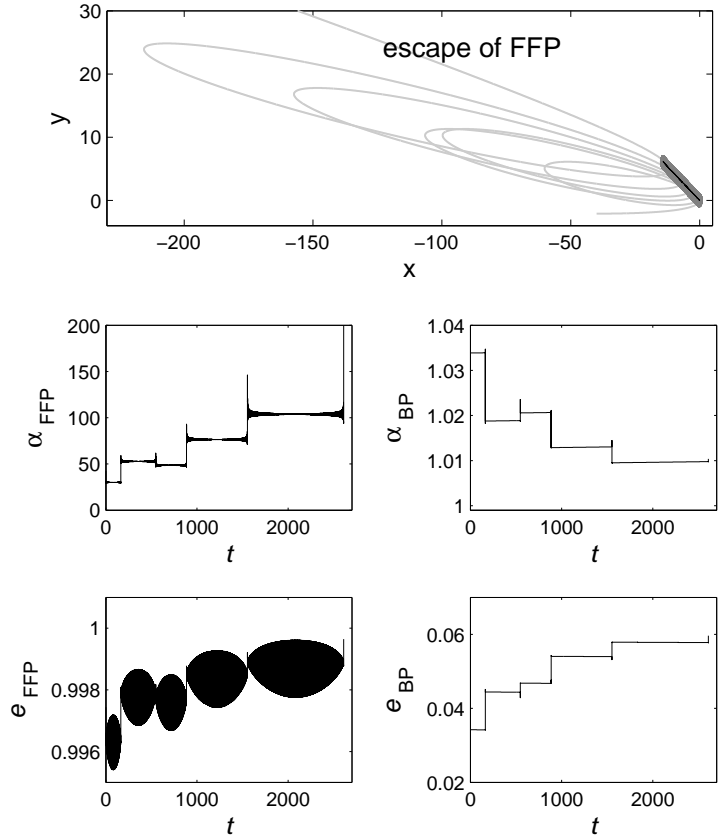}
\end{center}
\caption{(Top) An example of FFP escaping case following a period of temporary capture ({\it indirect - flyby}) (FFP: light gray, BP: dark gray, star: black). Corresponding evolution of stellarcentric semi-major axis (middle-row) and eccentricity (bottom-row) of the FFP and BP until the FFP's escape. We see that the planets jump from one orbital resonance to another (see Fig. 4 as well)}
\label{fig:01}       
\end{figure}

\begin{figure}
\begin{center}
  \includegraphics[width=11cm] {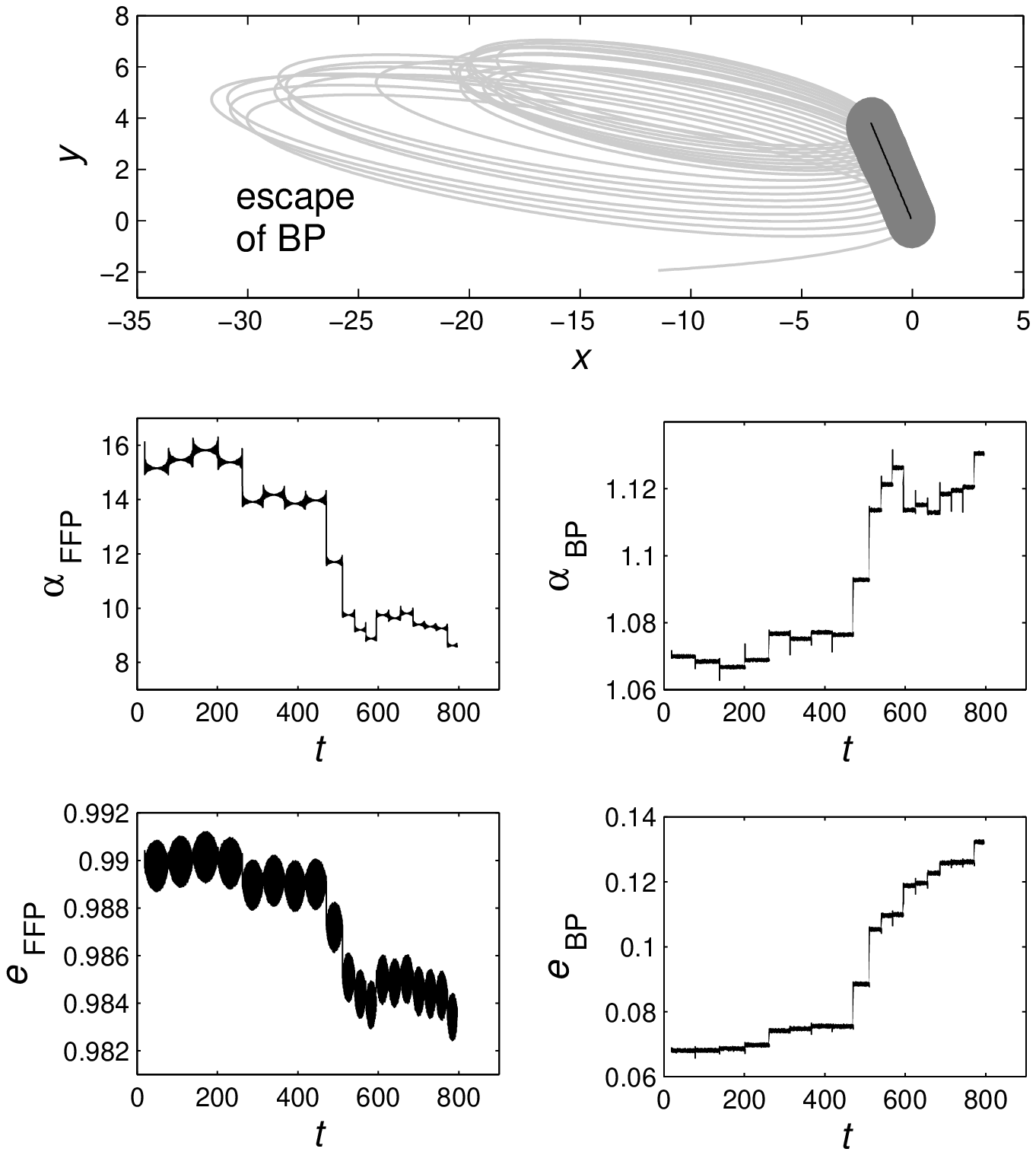}
\end{center}
\caption{(Top) An example of BP escaping case following a period of temporary capture of the FFP ({\it indirect - exchange}) (FFP: light gray, BP: dark gray, star: black). Corresponding evolution of stellarcentric semi-major axis  (middle-row) and eccentricity (bottom-row) of the FFP and BP until the BP's escape. We see that the escape of the BP is the final outcome of a sequence of resonances, which drive the FFP to lower values of ($a, e$), while the BP's elements increase. Note: The upper left corner panel
does not depict the escape of the BP, which happens at a much later time.}
\label{fig:03}       
\end{figure}

The case of capture of the FFP as a satellite of the BP $(E13<0)$,
although possible, never appeared in our numerical experiments.
The exit test is performed at time $t_{test} = 80 r_0 / 0.2236 v_0
\approx  360 r_0 / v_0 \approx 57.3 T$, where $T$ is the unperturbed
orbiting period of the BP. At this time the FFP, moving at its
initial velocity, would have traveled a distance twice its initial
distance from the star. Typical cases of FFP and BP escapes
(since we cannot be sure whether temporary capture is a {\it real} final
outcome, see Section \ref{sec:decay}) are given in Figs. \ref{fig:01} and \ref{fig:03}. It should be noted that {\it temporary capture} appears for both prograde and retrograde orbits (Gayon, 2009), which, as a rule, originate from corresponding initial positive or negative impact
parameters. However the opposite is not rare (initial negative (positive) impact parameter leading to prograde (retrograde) orbits). From this we see that the ``hidden"
physics below a final outcome (i.e. the exact path leading to a certain outcome) is not unique.

Finally it should be noted that an initial velocity of the FFP $10\%$ higher than the ``parabolic" one used in the paper (for which the total energy of the system remains still negative) results in a decrease of temporary captures.

\section{Scattering is fractal}

\subsection{Fractal properties of the initial-value space}
\begin{figure}
\begin{center}
  \includegraphics[width=11cm] {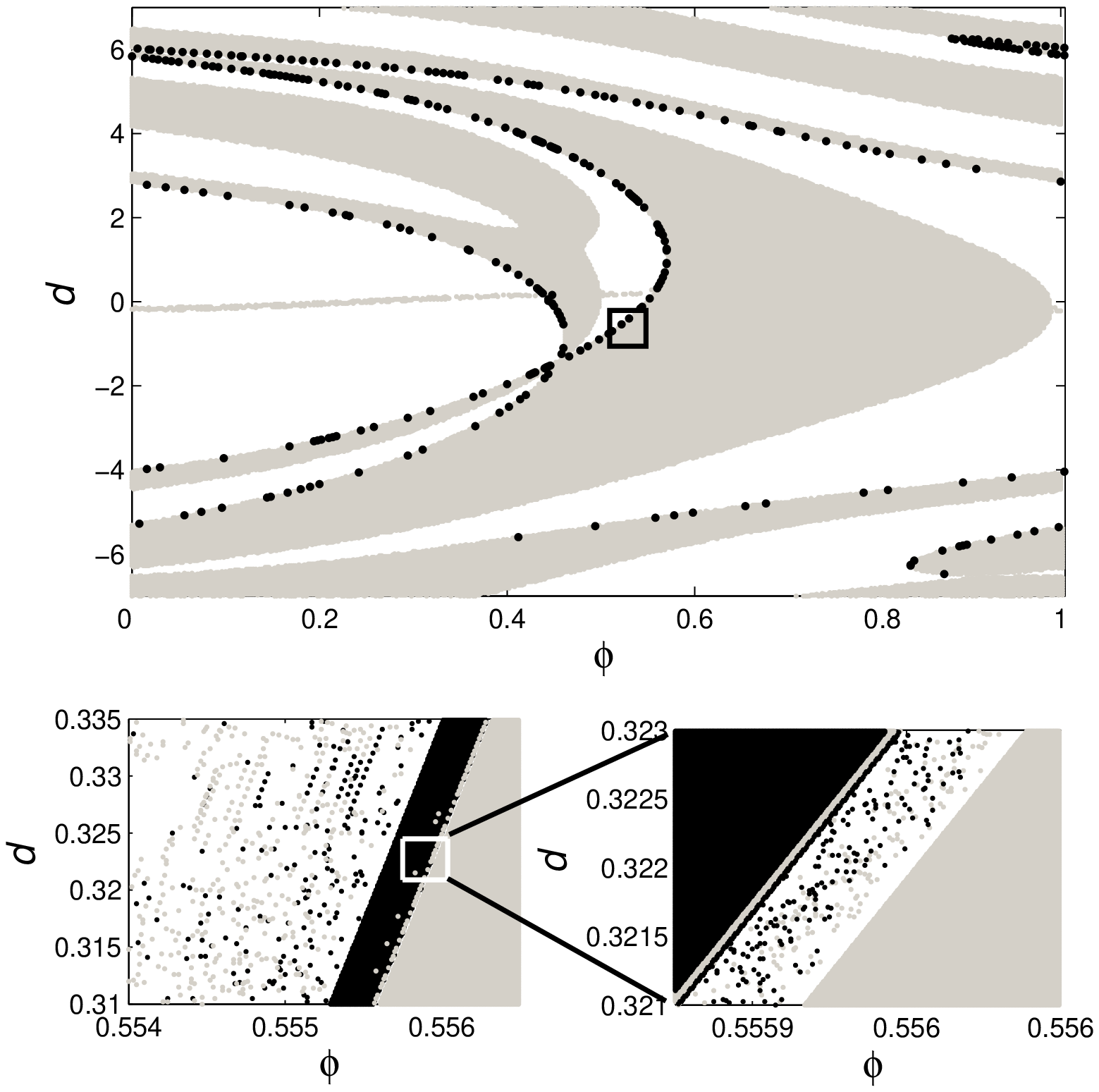}
\end{center}
\caption{(Top) Initial-value space for the scattering of a FFP with mass
$m = m_J$. Each initial condition $\phi, d$ is coded to represent
the state of the system after the scattering, at time $t_{test}=57.3T$.
By this time, the majority of initial conditions have led to either
flyby (gray) or temporary capture (white). Initial conditions corresponding to exchange (black)
constitute the boundary between the other two basins. Note that temporary capture (white) is not a definite outcome and the corresponding orbits may later on decay to either flyby or exchange.
(Bottom-left) Magnification of the region defined by the black box and (bottom-right) a further magnification of the sub-region contained in the white box. The occurrence of scattered initial condition leading to flyby and exchange within the basin of temporary capture (white), leads to great uncertainty in predicting the final state of the system for initial conditions close to the boundary. The two successive magnifications show an apparent self similar structure of the boundary, with fractal boundaries intertwined with smooth boundaries at finer and finer scales.}
\label{fig:2}       
\end{figure}

In Fig. \ref{fig:2} we present the final states corresponding to each initial condition in ($\phi, d$)
for the scattering of a FFP with mass equal to that of the BP
($m =m_J$).
Each initial condition pair ($\phi,d$) is coded
according to the final state of the system at the time of the exit
test, $t_{test} = 57.3 T$. Gray represents {\it flyby}, white
represents \textit{temporary capture} and black \textit{exchange}. As it is
illustrated in Fig. \ref{fig:2}, flyby and temporary capture of the
FFP are the most probable ($>99{\%}$) outcomes, while the
probability of exchange is much smaller ($\approx 0.1{\%}$). At this
resolution, it seems that there is a clear boundary between the
three basins of final states, with the basin of exchange
constituting the border between flyby and temporary capture.

Zooming twice in the region defined by the small black box in Fig.
\ref{fig:2}, we see that the boundary between the three basins is not smooth
(Fig. \ref{fig:2} bottom). Within the region leading to temporary capture (white),
there are scattered initial conditions that lead
to the other two final states, a fact repeated at smaller and
smaller scales. This indicates
uncertainty in predicting the final state of the system for regions
close to the boundary and a possible fractal structure. On the other
hand there are regions where the boundary between two basins appears
to be smooth i.e. the boundary between flyby (gray) and temporary capture
(white) basins in Fig. \ref{fig:2} bottom-right. The coexistence of fractal and
smooth (non fractal) boundaries intertwined in arbitrarily fine
scales is common in Hamiltonian systems (e.g. see Bleher et al.,
1990). From the two successive enlargements shown in Fig.
\ref{fig:2} it is also apparent that the non-smooth border looks
qualitatively self-similar for at least two scales of magnification.

\subsection{Estimating the uncertainty exponent}
In order to estimate the boundary (between basins) dimension, we apply the {\it final state
sensitivity method} introduced by Bleher et al. (1990). The basic idea behind this
method is to estimate the probability, $p(\varepsilon)$, to make a wrong prediction
of the final state when there is an uncertainty $\varepsilon$ in the
initial conditions. If the boundary is a smooth 1-D geometrical object (i.e. a line) this probability would be exactly proportional to $\varepsilon$. In a different case (i.e. a fractal boundary) we expect this probability to scale differently with $\varepsilon$. In any case, the scaling of $p$ with $\varepsilon$ gives the co-dimension of the boundary, $a$, from which the capacity dimension of the boundary follows as $d=D-a$, with $D$ the dimension of the embedding space (see Bleher et al. 1990). The method is the following:

In a box [$\phi, d$] on the initial-value space we select at random
25,000 ``central" initial conditions ($\phi, d$). Each ``central"
initial condition is perturbed along the $d$ axis by an amount
$\varepsilon$, so that we get two ``side" initial conditions ($\phi,
d + \varepsilon$) and ($\phi, d - \varepsilon$). If the three
initial conditions (the ``central" and the two ``side" ones) do not
all lead to the same final state, the ``central" condition is
considered as $\varepsilon$-{\it uncertain}. For every $\varepsilon$ we
calculate the fraction, $f(\varepsilon)$, of $\varepsilon$-uncertain
``central" initial conditions and plot it as a function of
$\varepsilon$ in log-log axes. The points ($\varepsilon ,
f(\varepsilon$) ) should lie on a straight line $\log f = a \log
\varepsilon + b$, where the slope, $a$, is the co-dimension of the
boundary, named in this case {\it uncertainty exponent}. The capacity dimension, $d$, of the boundary
is $d = D - a$, with $D=2$ (the dimension of the
initial-value space). A smooth boundary has $a = 1$,
so that $d = 1$ (a line). If $a < 1$ (as we find in most cases here), the boundary is fractal and $d
> 1$.

\begin{figure}
\begin{center}
  \includegraphics[width=11cm] {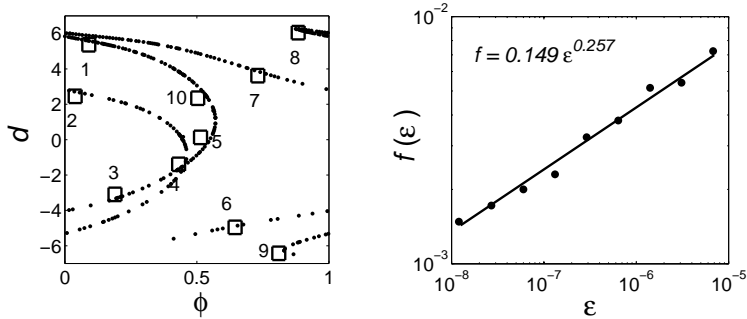}
\end{center}
\caption{(Left) Black boxes denote 10 regions of the initial
value-space for which the uncertainty exponent was estimated
(dimensions of the regions \textit{$\delta \phi $ }= 0.003 and
\textit{$\delta $d} = 0.06). (Right) Fraction of uncertain initial conditions, $f$, as function
of the uncertainty, $\varepsilon$, for region 3 and for the boundary of
flyby, considering the other two basins as one. The uncertainty exponent is estimated to
$\alpha = 0.257 \pm 0.014$, which gives a boundary dimension of $d \approx 1.743$.}
\label{fig:4}       
\end{figure}

To estimate the uncertainty dimension we select 10 sub-regions of
the initial-value space, shown in Fig. \ref{fig:4}. In each region we compute
the capacity dimension of the boundaries of the basins of (i)
exchange and (ii) flyby, considering in each case the remaining two
basins as a single one. The scaling of $f$ with
$\varepsilon$ for region 3 of Fig. \ref{fig:4} (left) and for the basin of flyby
is shown in Fig. \ref{fig:4} (right). For $\varepsilon < 10^{-5}$ the function $f(\varepsilon)$
is very well fitted by a straight line for at least 3 orders of
magnitude of $\varepsilon$, which is a strong indication that our
results show a real fractality of the boundary and are not an
artifact of the method. The fit gives $a = 0.257 \pm 0.014$ and, therefore, $d \simeq 1.743$. The corresponding values
of the uncertainty exponent in the other regions of Fig. 6 range in the interval $0.209 < a < 0.346$.

It should be noted that the uncertainty exponent, $a$, is a decreasing function of the test time, $t_{test}$. This is because, as we will see in more detail  below, temporary captures decay over time in an exponential fashion. Thus, on much longer time scales, there appear to be only two possible outcomes - instead of three - and the boundary  becomes smoother.  However, as we will see in the following, the number of temporary captures does not seem to go to zero on reasonably long time scales (see section \ref{sec:decay}).

\section{Final state statistics}

\subsection{Probabilities of exchange, flyby and temporary capture}
We have performed a large number of numerical experiments for various values of the ratio $m/m_J$ and in this section we give the statistics at time $t_{test} = 57.3T$, concerning (direct) exchange, (direct)
flyby and temporary capture as functions of $m/m_J$.

\begin{figure}
\begin{center}
  \includegraphics[width=11cm] {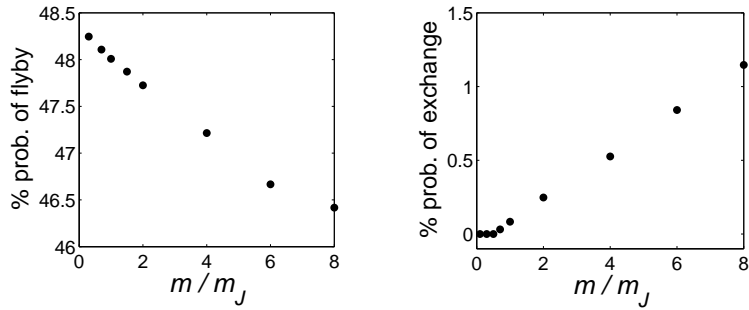}
\end{center}
\caption{Fraction of orbits that have resulted in flyby and exchange by the time
of the exit test ($t_{test}= 57.3T$) as functions of the FFP's mass. The third outcome, namely temporary capture of the FFP, is the supplement of the other two sets and is an increasing function of the FFP's mass in this range of $m$. Note that the probability of exchange falls to zero for $m<0.5m_J$ (see section \ref{sec:44}). }
\label{fig:5}       
\end{figure}

The probabilities of exchange and flyby as functions of
the FFP's mass are shown in Fig. \ref{fig:5}. Each point in the graphs represent a simulation of 350,000 initial conditions. Roughly half of the cases end up in direct flyby, while a small percentage (of the order of 1\%) ends up in direct exchange. The rest of the cases ($\approx50\%$) correspond to temporary capture of the FFP. It is interesting to note that, for a relatively large interval, the exchange and flyby probabilities depend linearly on the FFP's mass. Another interesting point is that there is a lower mass limit ($m=0.5m_{J}$), below which direct exchange cannot occur.
An explanation of this fact is given in section (\ref{sec:44}). A similar result concerning the lower mass limit for exchange has been reported by Boyd and MacMillan (1993)
in the case of scattering of a star from a binary star of equal mass members,
within a configuration similar to the one used here. In particular, these authors found
that the probability of exchange drops to zero for masses of the incoming body between 0.3 and 0.8 $m_{star}$ ($M$). For the problem of equal binary masses the probability of exchange increases rapidly with the mass of the incoming body, since in this case either one of the binary members can escape. In our case, the masses of the BP and the FFP are much smaller than that of the star and
the probabilities of exchange, flyby and temporary capture do not vary strongly with the mass of the FFP.

\subsection{Orbital distribution in temporary capture}
\begin{figure}
\begin{center}
  \includegraphics[width=11cm] {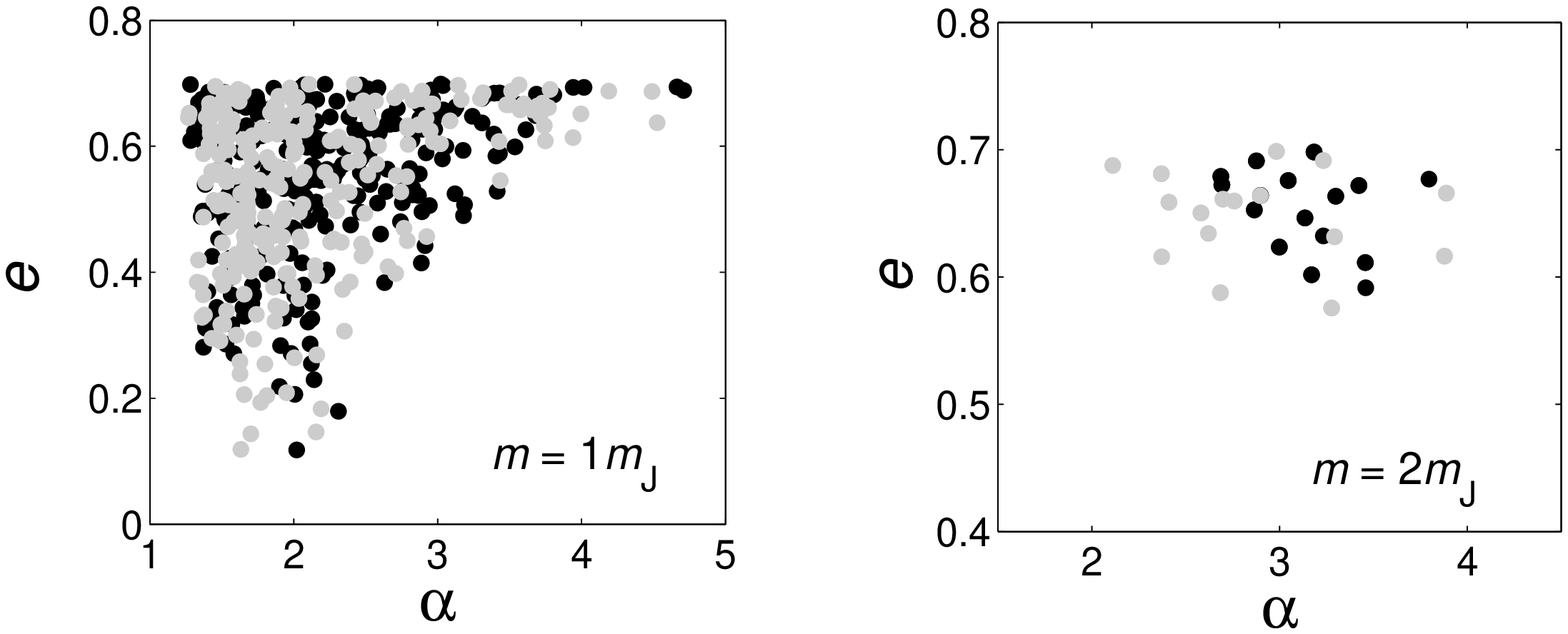}
\end{center}
\caption {Stellarcentric orbital elements of the FFP (black) and the BP (gray) in temporary capture, for orbits for which both planets have moderate values of semi-major axis and eccentricity ($\alpha<50r_0$, $e<0.7$). The elements are computed at time $t_{test}=57.3T$.}
\label{fig:6}       
\end{figure}

\begin{figure}
\begin{center}
  \includegraphics[width=11cm] {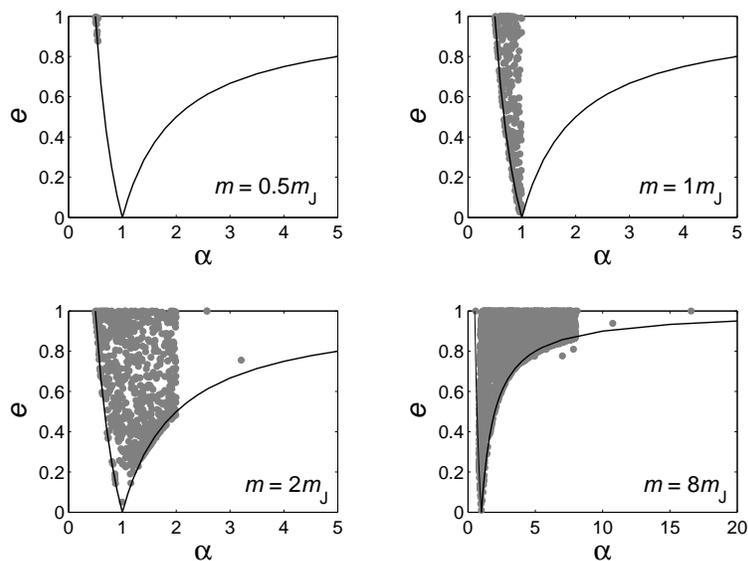}
\end{center}
\caption {Eccentricity ($e$) as a function of the semi-major axis ($\alpha$)
of the FFP in the event of direct exchange, for four different values of the FFP's mass:
$m=0.5m_{J}$, $m=1m_{J}$, $m=2m_{J}$ and $m=8m_{J}$. The stellarcentric osculating values of $\alpha$ and $e$ are computed at time $t_{test}=57.3 T$ for the orbits that have resulted in exchange by that time. The solid lines are constant pericenter and apocenter curves $e=1-1/a$ for $a>1$ and $e=1/a-1$ for $a<1$).}
\label{fig:7}       
\end{figure}

In temporary captures the orbital elements of both BP and FFP vary within a wide range.
Shortly after scattering, the majority of captured FFPs have very elongated orbits ($e>0.99$) with semi-major axis $1.998 r_0 < a <3.3\cdot10^7r_0$, while the BP is slightly perturbed form its initial orbit, a result corroborated by Perets and Kouwenhoven (2012). In 90\% of the cases the orbits of the two planets do intersect. However, there are also a few cases where both planets have moderate values of $a$ and $e$. Fig. \ref{fig:6} shows such cases for two values of the FFP's mass. Depending on the mass, these orbits make up from 0.01{\%} ($m=2m_J$) to 0.2{\%} ($m=0.5m_J$) of the initial sample of orbits. Finally, although the eccentricity and semi-major axis show important variations, there is an approximate conservation of pericenter distance for both bodies, with $q \approx 1.0075$ for the FFP and $q \approx 1.1057$ for the BP.

As time passes ($t=600,000 T$) the picture changes, since now 75\% of the orbits do not intersect, as most of the interacting orbits have decayed to flyby or exchange. In all these orbits the FFP has pericenter distance in the range $2 r_0 < q <1291 r_0$, while the BP has semi-major axis in the range $0.9 r_0 < a < 1.2 r_0$ and moderate eccentricity. There are also a few orbits where the BP and FFP exchange positions (the FFP jumps to an inner orbit and the BP goes outwards). As we show in section \ref{sec:decay}, there exist bound orbits that survive for very long times.

It should be noted that tidal interactions between the FFP and the star, in the case of extended bodies, are expected to stabilize the FFP's orbit, if the percenter distance, $q$, is relatively short. Due to this process, named {\it tidal circularization}, the eccentricity of a planet is decreased through the combined effects of energy dissipation, through tidal friction, and conservation of angular momentum. From analysis of our results we have found that, after $t = 200,000 T$, 6\% of the orbits have $q < 0.1$ AU; this fraction drops to 2\% after $t=600,000 T$. The fraction of orbits with small pericenter distance is of comparable order of magnitude to the long-time surviving orbits found in section \ref{sec:decay} and, therefore, are not negligible in the final analysis, since they might increase the fraction of FFPs eventually captured.

\subsection{$e$ vs. $a$ in direct exchange}
\label{sec:44}
Fig. \ref{fig:7} shows an interesting ``regularity" in the event of exchange.
The eccentricity of the captured FFP is plotted as a function of
its semi-major axis, for various values of the FFP's mass. In all cases
$a$ and $e$ are contained in a region in $a, e$ space bounded by two curves, $e=1-(1/a)$ for $a>1$
and $e=(1/a)-1$ for $a<1$. The first one corresponds to constant pericenter distance $q=1.0$ and the second one to constant apocenter distance, $a(1+e)=1$. By setting $e=1$ in the second equation we obtain a lower bound for the semi-major axis of the FFP, which
equals to $a=0.5r_{0}$, independent of the FFP's mass. There is also
a less clear upper bound of the value of the semi-major axis, which
seems to depend on the FFP's mass. This upper bound is a result of energy conservation.
In particular, the energy taken by the escaping BP is gained at the expense of the star-FFP energy.
Since the binding energy of the initial star-BP pair is
$E_{J}=-(GMm_{J})/2 r_{0}$ and the FFP has initially
zero binding energy, the binding energy of the new pair will be
equal to or less than $E_{J}$, so that $ -(\textit{GMm})/2a  \le
-(\textit{GMm}_{J})$/2$r_{0}$. It follows that the
semi-major axis of the FFP will be $a<(m/m_{J})r_{0}$. According to this, an FFP with $m=m_{J}$ will
have at most $a=1r_{0}$, which is apparently the case in Fig. \ref{fig:7}.
Now consider an FFP of mass $m<0.5m_{J}$. This will have semi-major axis $a<0.5r_{0}$.
However this value is below the lower bound $a=0.5r_{0}$ (imposed by
the lower bounding curve ($e=(1/a)-1$)). Therefore a direct exchange cannot occur for
masses of the FFP below $0.5m_{J}$.

\begin{figure}
\begin{center}
  \includegraphics[width=11cm] {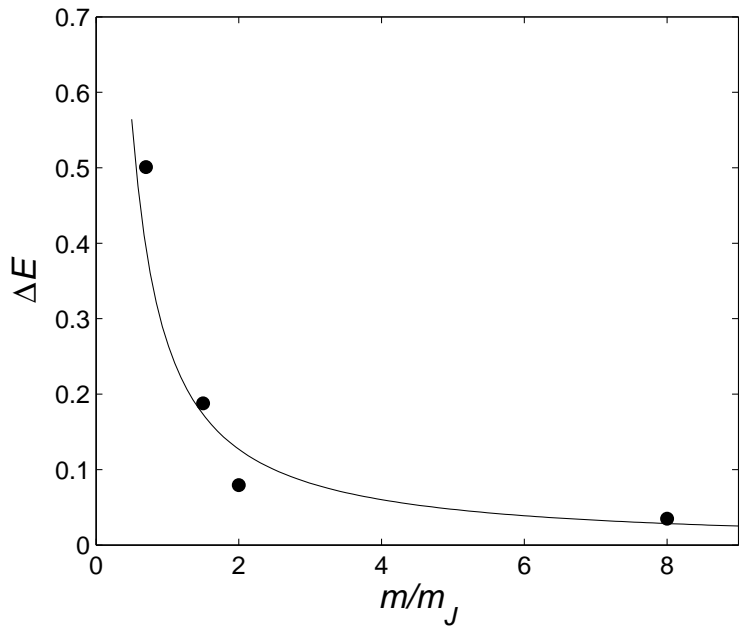}
\end{center}
\caption {Energy loss (per unit mass) from the FFP-star pair as a function of the FFP's mass. We see that a curve of the form $\Delta E \approx 1/m$ is fitting the data with a coefficient of determination $R^2 \simeq 0.9$. $\Delta E$ is computed for the orbits with semi-major axis that exceeds the upper bound $a < m/m_J)r_0$. $\Delta E$ is the difference between the true binding energy of the FFP and the energy that corresponds to the upper bound.}
\label{fig:DEexchange}       
\end{figure}

 An interesting point, coming out from Fig. 9, is that in some direct exchange outcomes for $m/m_J = 2$ and $m/m_J = 8$ we observe that $a > (m/m_J) r_0$. We have looked in detail onto this and we have found that the additional energy $\Delta E$ of the FFP-star pair is taken from the FFP-BP one, which is non-negligible when the FFP is relatively massive. In Fig. \ref{fig:DEexchange} we see that $\Delta E$ obeys an approximate relation of the form $\Delta E \propto 1/m$.

\section{Exponential decay of temporary captures}\label{sec:decay}
\begin{figure}
\begin{center}
  \includegraphics[width=11cm] {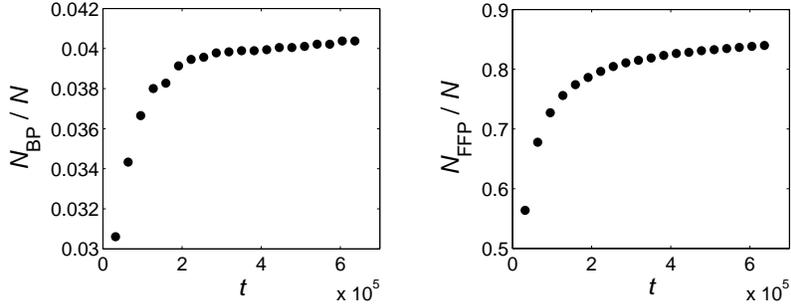}
\end{center}
\caption{(Left) Fraction of orbits that decay to exchange vs. time. (Right) Fraction
of orbits that decay to flyby vs. time. $N = 6,500$ is the initial sample of orbits that were in temporary capture at time $t_{test}=57.3T$
and were integrated up to time $t=600,000T$.}
\label{fig:8a}       
\end{figure}

\begin{figure}
\begin{center}
  \includegraphics[width=11cm] {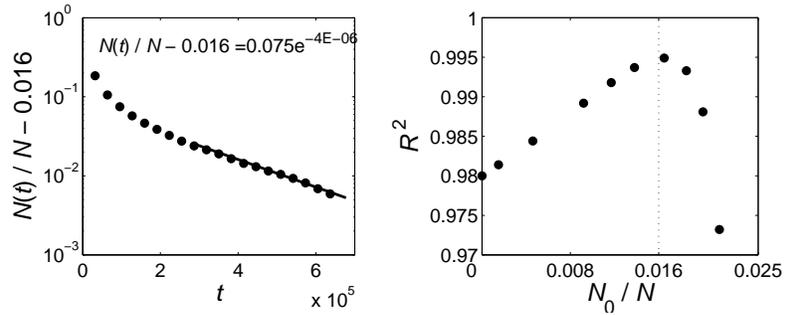}
\end{center}
\caption{(Left) Fraction of orbits that remain in temporary capture until $t=600,000T$ and the curve that fits best the data.
(Right) Coefficient of determination, $R^{2}$, as a function of $N_{0}$/$N$. The best fit occurs for $N_{0}$/$N=0.016$, where $N_{0}$/$N$ is the fraction of orbits that are expected to remain in capture as $t\rightarrow\infty$. $N = 6,500$ is the initial sample of orbits that were in temporary capture at time $t_{test}=57.3T$ and were integrated up to time $t=600,000T$.}
\label{fig:8}       
\end{figure}

A temporary capture configuration at $t_{test}$ can decay through the
escape of either the BP or the FFP. In order to study the decay
process, we continued the numerical integration of 6,500 cases that
were in temporary capture at $t_{test}$ = 57.3$T$ up to $t_{final} = 600,000 T$. Fig. \ref{fig:8a} shows the fraction
of orbits that have resulted in indirect exchange (left) and flyby (right) as functions of time. We see that more than $90${\%} of the orbits, that were in temporary capture at time $t_{test}$,
will result in flyby and approximately 4{\%} of them will result in
exchange as $t \to \infty$. Fig. \ref{fig:8}
shows a lin-log plot of the fraction $N(t)$/$N$ of orbits, that are still in
temporary capture after time $t$, vs. $t$. We see that, after $t =
200,000 T$, there seems to be an exponential fall-off of this
fraction. By fitting a least square curve of the type $N(t)/N - N_0/N = \beta \cdot
e^{- \lambda t}$ to the data we find (through a $\chi^2$ test, see Fig. \ref{fig:8}) that the best fit is:
$N(t)/N - 0.016 = \beta \cdot e^{- 4\cdot10^{-6}t}$. This means that there is a
fraction of orbits, $N_0/N \simeq 0.016$, that are expected to remain in
temporary capture for very long times. It should be noted that the fractal properties of the basin boundaries persist, even after the decay of the temporary capture configurations to either indirect flybys or indirect exchanges.

\section{Prograde and retrograde orbits}

\begin{figure}
\begin{center}
  \includegraphics[width=5cm] {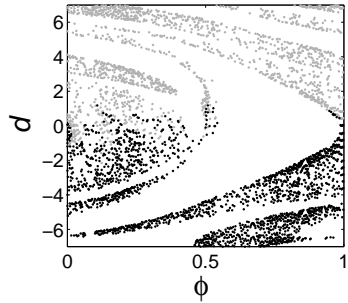}
\end{center}
\caption{Initial conditions
corresponding to orbits that remain in temporary capture until $t=60,000$ $T$. Grey indicates prograde orbits and black indicates
retrograde ones.}
\label{fig:8b}       
\end{figure}

Fig. \ref{fig:8b} is a plot of the initial ($\phi$, $d$) values that correspond to prograde (gray) and retrograde (black) orbits of the two planets, that remain bound up to $t=60,000 T$. As expected, orbits with positive impact parameter end up mostly as prograde and those with negative impact parameter mostly as retrograde. However, there are orbits for which the FFP or the BP change their initial revolution direction. This happens for low values of the impact parameter, between $-2r_{0}$ and 2$r_{0}$ for which the FFP's orbit passes very close by the BP's orbit (see Fig. \ref{fig:8b} where we have prograde orbits (gray) with $d<0$ and retrograde orbits (black) with $d>0$). In the majority of the cases it is the FFP that changes its revolution direction. This is to be expected, since the FFP orbit is nearly parabolic and therefore can easily become a very eccentric ellipse. The angular momentum exchange between the FFP and the BP depends on the minimal distance and relative phase of the two bodies, during their encounter. Thus, the smaller the impact parameter, the closer the two bodies can get. This increases the magnitude of the angular momentum exchanged and - depending on the relative phase - increases the probability that the FFP's orbit will change its revolution direction around the host star.

\section{Summary and conclusions}
In the present paper we have studied the gravitational scattering of a FFP by a star-planet pair and we have shown that it is chaotic, in the sense that the boundary of a region in initial conditions that leads to a specific outcome (i.e. flyby, capture or exchange) is fractal. In this way our statistical approach is fully justified.

One interesting result, coming out from our analysis is that the probability of exchange, flyby or temporary capture is, for values of the FFP's mass spanning one order of magnitude, a linear function of
the FFP's mass, decreasing in the case of flyby and increasing in the cases of exchange and temporary capture. It should be noted that for masses of the FFP below $0.5m_J$ exchange is not possible, due to energy conservation.

Another interesting result is that, in the case of exchange or temporary capture, FFPs can be captured in orbits with {\it moderate} values of ($a, e$), with the simultaneous excitation of the BP's orbital elements. In the case of temporary capture, orbits where both FFP and BP have moderate values of ($a, e$) make up $\approx 0.2{\%}$ of all captured orbits (depending on the FFP's mass). Restricting ourselves to non-Hill stable interactions, there is a
cumulative 4{\%} probability that a FFP, passing within a distance of $7r_0$ from the star (where $r_0$ is the radius of the BP's circular orbit) will be captured in orbit around the star. This cumulative probability is the sum of direct exchange (0.2{\%}), indirect exchange (2{\%}) and asymptotic captures (1.6{\%}). According to Donnison (2008), the extension to 3D should increase the probability of capture of the FFP by the star-BP pair.

The scattering induces an excitation of the orbital elements of the final system in both exchange and temporary capture. In exchange, the FFP can be captured in highly eccentric orbits ($e>0.9$) with semi-major axis up to $a = (m/m_{J})r_{o}$ (where $m$ the mass of the FFP and $m_J$, $r_0$ the mass and semi-major axis of the bound planet). In temporary capture, the majority of the FFP orbits that survive up to $t\approx600,000$ orbital periods of the BP have very large and elongated orbits ($e>0.97$ and $a > 5.000r_0$) with the BP being only slightly perturbed from its initial orbit. Nevertheless 75\% of these orbits do not intersect.

The above results come out from a model with two main simplifications: (i) co-planar configuration and (ii) absence of dissipation and some minor simplifications: (iii) circular orbit of the BP, (iv) fixed planet to star mass ratio and (v) parabolic orbit of the FFP. It would be of great interest to perform further ``experiments", by introducing (i) the third dimension and (ii) a mechanism of tidal dissipation acting {\it after} the capture, in order to study the long-time evolution of FFP's in high-$e$ and high-$a$ orbits.

{\em Acknowledgements. We would like to acknowledge two anonymous referees for their useful comments. VS would like to thank S. Sgardeli and L. Christodoulou for constructive discussions and comments on the manuscript.}

\end{document}